\documentclass[showemail,pre]{revtex4}

\usepackage{graphicx}   
\usepackage{bm}
\begin{document}

\title{
Flicker Noise in a Model of Coevolving Biological Populations
}
\author{Per Arne Rikvold$^{1,2,}$}\email{rikvold@csit.fsu.edu}
\author{R.K.P.\ Zia$^{2,}$}\email{rkpzia@vt.edu}
\affiliation{
$^1$School of Computational Science and Information Technology,
Center for Materials Research and Technology, and Department of Physics,
Florida State University, Tallahassee, Florida 32306-4120\\
$^2$Center for Stochastic Processes in Science and Engineering,
Department of Physics, Virginia Polytechnic Institute and State University,
Blacksburg, VA 24061-0435
}
\date{\today}

\begin{abstract}
We present long Monte Carlo simulations of a simple model of biological
macroevolution in which births, deaths, and mutational changes in the
genome take place at the level of individual organisms. The model
displays punctuated equilibria and flicker noise with a $1/f$-like power
spectrum, consistent with some current theories of evolutionary
dynamics. 
\end{abstract}

\maketitle

\section{Introduction}
\label{sec:I}

The long-time dynamics of
biological evolution have recently attracted considerable interest
among statistical physicists \cite{DROS01}, who find in this field new
and challenging interacting nonequilibrium systems.
An example is the Bak-Sneppen model
\cite{BAK93}, in which interacting species are the basic units, and less
``fit" species change by ``mutations" that trigger avalanches that may
lead to a self-organized critical state. However, in reality both mutations
and natural selection act on {\it individual organisms\/}, and it is
desirable to develop and study models in which this is the case. 
One such model was recently introduced by Hall, Christensen, and
coworkers \cite{HALL02,CHRI02}. To enable very long Monte Carlo (MC)
simulations of the evolutionary behavior,
we have developed a simplified version of this model, for
which we here present preliminary results.

\section{Model and Numerical Results}
\label{sec:M}

The model consists of a population of individuals with a haploid genome of $L$
binary genes \cite{EIGE71,EIGE88}, so that the total number of potential
genomes is $2^L$. The short genomes we have been able to study
numerically (here, $L=13$) should be seen as coarse-grained
representations of the full genome. We thus consider each
different bit string as a separate ``species" in the rather loose sense
that this term is used about haploid organisms. 
In our simplified model the population evolves asexually in discrete,
nonoverlapping generations, and
the population of species $I$ in generation $t$ is $n_I(t)$.
The total population is $N_{\rm tot}(t) = \sum_I n_I(t)$.
In each generation, the probability that an individual of species $I$ has
$F$ offspring before it dies is $P_I(\{n_J(t)\})$, while it
dies without offspring with probability $1-P_I$.
The reproduction probability $P_I$ is given by 
\begin{equation}
P_I(\{n_J(t)\})
=
\frac{1}{1 + \exp\left[ - \sum_J M_{IJ} n_J(t)/N_{\rm tot}(t)
+ N_{\rm tot}(t)/N_0 \right]}
\;.
\label{eq:P}
\end{equation}
The Verhulst factor $N_0$ \cite{VERH1838}, which
prevents $N_{\rm tot}$ from diverging,
represents an environmental ``carrying capacity'' due to limited
shared resources. The time-independent
interaction matrix $\bf M$ expresses pair interactions between different
species such that the element $M_{IJ}$ gives the effect of the population
density of species $J$ on species $I$. Elements
$M_{IJ}$ and $M_{JI}$ both positive represent symbiosis or mutualism,
$M_{IJ}$ and $M_{JI}$ both negative represent competition, while
$M_{IJ}$ and $M_{JI}$ of opposite signs represent predator-prey relationships.
To concentrate on the effects of interspecies interactions,
we follow \cite{HALL02,CHRI02} in taking $M_{II} = 0$.
As in \cite{CHRI02}, the offdiagonal elements of $M_{IJ}$ are randomly
and uniformly distributed on $[-1,1]$.
In each generation, the genomes of the individual offspring organisms 
undergo mutation with probability $\mu / L$ per gene and individual.

MC simulations were performed with the following
parameters: mutation rate $\mu = 10^{-3}$ per individual,
carrying capacity $N_0 = 2000$, fecundity $F=4$, and genome length $L=13$.
For a system with ${\bf M} = {\bf 0}$ or only a single species and $\mu
= 0$, the steady-state total population is found by linear stability
analysis \cite{RIKV03} to be $N_0 \ln(F-1) \approx 2200$.
In this regime both the number of populated species and the
total population $N_{\rm tot}(t)$ are 
smaller than the number of possible species, $2^L = 8192$. This appears
biologically reasonable in view of the enormous number of different
possible genomes in nature.

An important quantity is the diversity of the population, which is
defined as the number of species with significant populations.
Operationally we define it as
$D(t) = \exp \left[S \left( \{ n_I(t) \} \right) \right]$,
where $S$ is the information-theoretical entropy (known in ecology as
the Shannon-Weaver index 
\cite{SHAN49}),
$ S\left( \{ n_I(t) \} \right)
=
- \sum_{\{I | n_I(t) > 0 \}} \left[ {n_I(t)}/{N_{\rm tot}(t)} \right]
\ln \left[ {n_I(t)}/{N_{\rm tot}(t)} \right] $.

\begin{figure}[t]
\includegraphics[width=.48\textwidth]{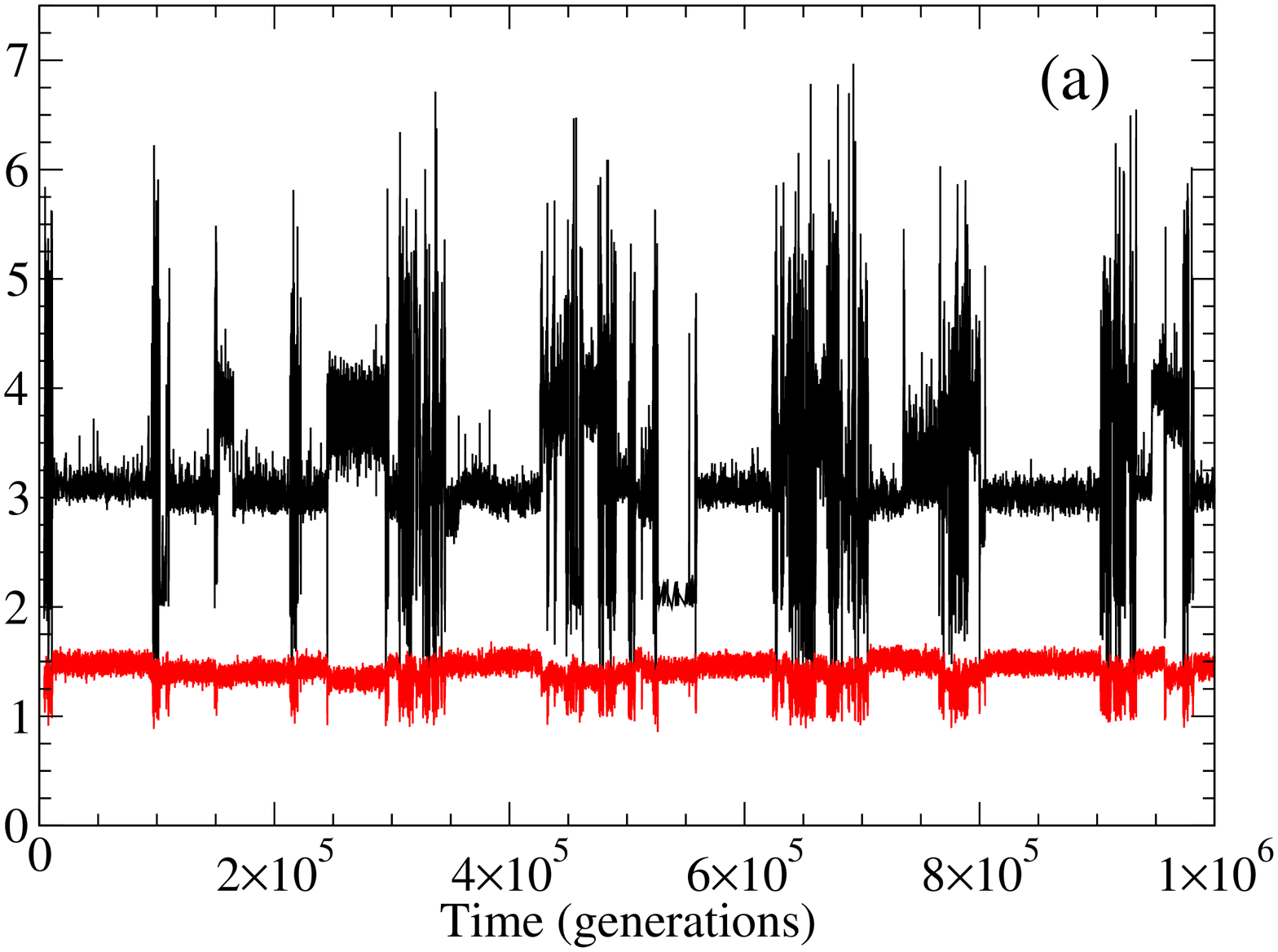}
\includegraphics[width=.48\textwidth]{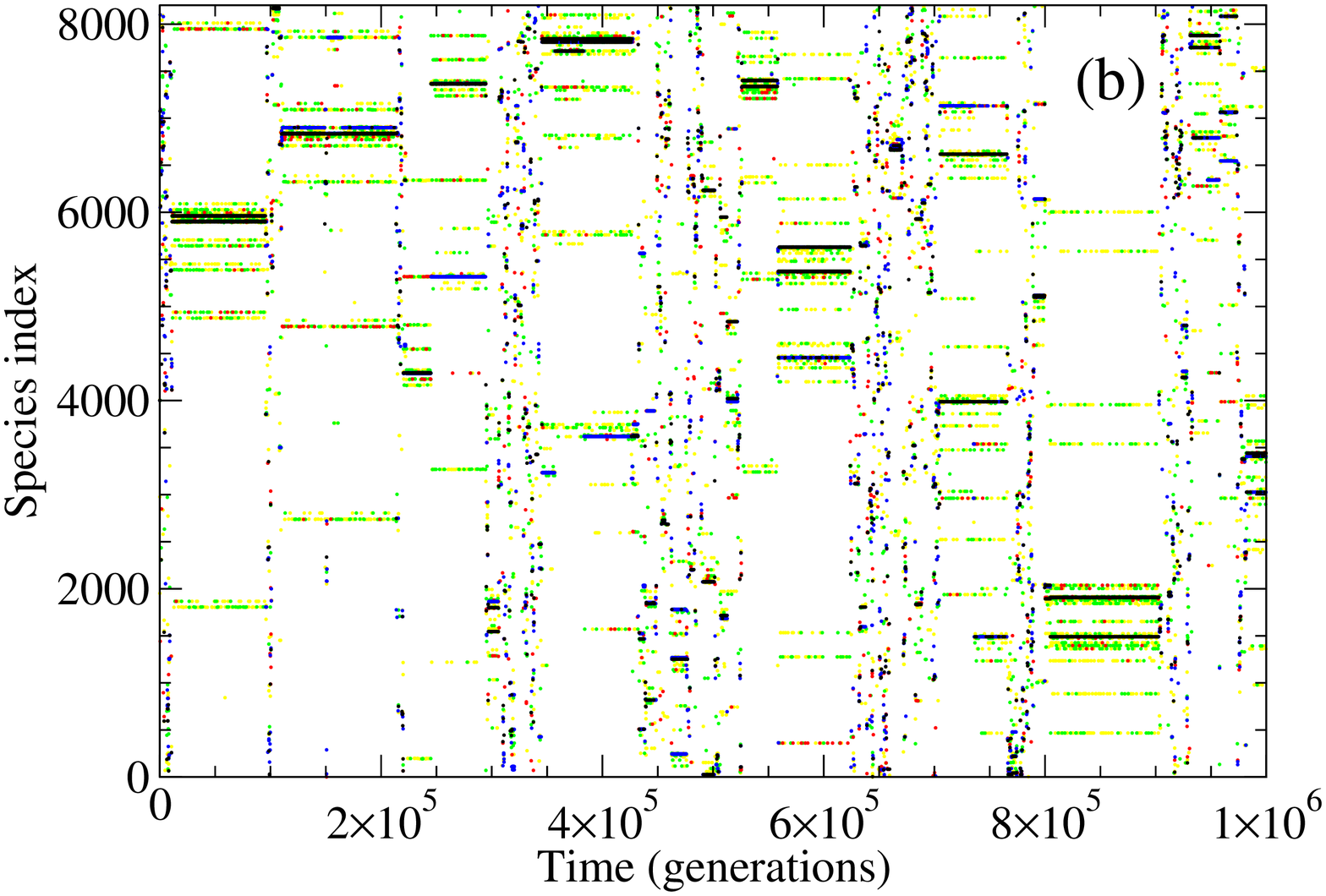}
\caption[]{
Results of a simulation of 10$^6$ generations with the parameters given
in the text.
({\bf a})
Time series showing the diversity, $D(t)$ ({\it black\/}), 
and the normalized total
population, $N_{\rm tot}(t)/[N_0 \ln (F-1)]$ ({\it red\/}).
({\bf b})
Species index $I$ vs time. 
The symbols indicate $n_I > 1000$ ({\it black\/}), 
$n_I \in [101,1000]$ ({\it blue\/}), $n_I \in [11,100]$ ({\it red\/}), 
$n_I \in [2,10]$ ({\it green\/}), and $n_I=1$ ({\it yellow\/}). 
}
\label{fig:fig1}
\end{figure}

Results for a run of $10^6$ generations are shown in
Fig.~\ref{fig:fig1}. In Fig.~\ref{fig:fig1}({\bf a}) are shown time
series of $D(t)$ and $N_{\rm tot}(t)$. We see relatively
quiet periods (quasi-steady states, QSS) punctuated by periods of high
activity. During the active periods the diversity fluctuates wildly,
while the total population falls below its typical QSS value. A
corresponding picture of the species index (the decimal
representation of the binary genome) is shown in Fig.~\ref{fig:fig1}({\bf b}), 
with grayscale indicating
$n_I(t)$. Comparison of the two parts of Fig.~\ref{fig:fig1} show 
that the QSS correspond to
periods during which the population is dominated by a relatively small
number of species, while the active periods correspond to transitions
during which the system is ``searching for" a new QSS. 

\begin{figure}[t]
\centering
\vspace{0.5truecm}
\includegraphics[width=.49\textwidth]{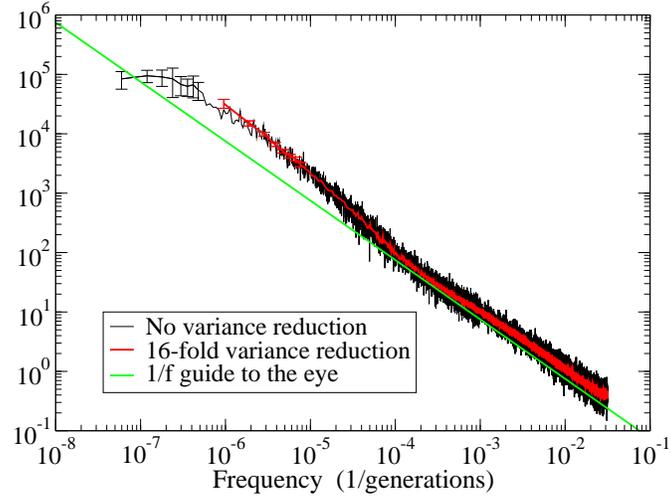}
\caption[]{
PSD of the diversity, based on nine independent simulations of $2^{25}$
generations each. The model parameters are those given in the text and used in 
Fig.~\ref{fig:fig1}. The $1/f$ like spectrum is indicative of very long-time
correlations and a wide distribution of QSS lifetimes. 
}
\label{fig:fig2}
\end{figure}

Closer inspection of Fig.~\ref{fig:fig1} suggests that there are shorter
QSS within some of the periods of high activity. This led us to consider
the power-spectral densities (PSD) of the diversity and total
population, measured in very long simulations of $2^{25} =
33\,554\,432$ generations. The PSD of the diversity
is shown in Fig.~\ref{fig:fig2} and indicates
that the model exhibits flicker noise with a spectrum near $1/f$ 
\cite{MARI83,MILO02} over at least four to five decades in frequency.

\section{Relevance to Evolutionary Biology}
\label{sec:C}

It has been much discussed in evolutionary biology whether species
evolve gradually or in a succession of QSS, punctuated by periods of
rapid change. 
The latter mode has been termed ``punctuated equilibria" by Gould and
Eldredge 
\cite{GOUL77,NEWM85}. 
There is also some indication that flicker noise is found in the fossil record
of extinctions, but due to the sparseness of the fossil evidence this is
a contested issue \cite{HALL96,NEWM99}. 
The model discussed here can at best be
applied to the evolution of asexual, haploid organisms such as bacteria,
and one should also note that no specific, biologically relevant information 
has been included in the interaction matrix. 
Nevertheless, we find it encouraging that such
a simple model of macroevolution with individual-based births, deaths, and
mutations can produce punctuated equilibria and flicker noise
reminiscent of current theories of biological macroevolution.

\section*{Acknowledgments}

We thank B.~Schmittmann and U.~T{\"a}uber for useful discussions, and 
P.A.R.\ thanks the Department of Physics, Virginia
Polytechnic Institute and State University, for its hospitality.
This research was supported by U.S.\ National Science Foundation Grant
Nos.\ DMR-9981815, DMR-0088451, DMR-0120310, and DMR-0240078,
and by Florida State University through the School of Computational
Science and Information Technology and the Center for Materials Research
and Technology.

%
%
%

\end{document}